# Routing Algorithms for Wireless Sensor Networks

**Ioan Raicu**
Department of Computer Science
Wayne State University
Detroit, MI 48202
iraicu@cs.wayne.edu

**Abstract**

Our contribution in this paper is e3D, a diffusion based routing protocol that prolongs the system lifetime, evenly distributes the power dissipation throughout the network, and incurs minimal overhead for synchronizing communication. We compare e3D with other algorithms in terms of system lifetime, power dissipation distribution, cost of synchronization, and simplicity of the algorithm.

## 1. INTRODUCTION

In this paper, we attempt to overcome limitations of the wireless sensor networks such as: limited energy resources, varying energy consumption based on location, high cost of transmission, and limited processing capabilities. Besides maximizing the lifetime of the sensor nodes, it is preferable to distribute the energy dissipated throughout the wireless sensor network in order to minimize maintenance and maximize overall system performance. For more in depth understanding of the problem statement and proposed algorithm, we refer the reader to the full length version of this paper [1].

Any communication protocol that involves synchronization between peer nodes incurs some overhead of setting up the communication. We attempt to calculate this overhead and to come to a conclusion whether the benefits of more complex routing algorithms overshadow the extra control messages each node needs to communicate. Obviously, each node could make the most informed decision regarding its communication options if they had complete knowledge of the entire network topology and power levels of all the nodes in the network. This indeed proves to yield the best performance if the synchronization messages are not taken into account. However, since all the nodes would always have to know everything, it should be obvious that there will be many more synchronization messages than data messages, and therefore ideal case algorithms are not feasible in a system where communication is very expensive. For both the diffusion and clustering algorithms, we will analyze both realistic and optimum schemes in order to gain more insight in the properties of both approaches. The benefit of introducing these ideal algorithms is to show the upper bound on performance at the cost of an astronomical prohibitive synchronization costs.

## 2. e3D: EXPERIMENTAL RESULTS

We introduce a new algorithm, e3D, and compare it to two other algorithms, namely directed, and random clustering communication. We take into account the setup costs and analyze the energy-efficiency and the useful lifetime of the system. We compare the algorithms in terms of system lifetime, power dissipation distribution, cost of synchronization, and simplicity of the algorithm. Our simulation results show that e3D performs only slightly worse than its optimal counterpart while having much less overhead. Therefore, our contribution is a diffusion based routing protocol that prolongs the system lifetime, evenly distributes the power dissipation throughout the network, and incurs minimal overhead for synchronizing communication. In our simulation, we use a data collection problem in which the system is driven by rounds of communication, and each sensor node has a packet to send to the distant base station. The diffusion algorithm is based on location, power levels, and load on the node, and its goal is to distribute the power consumption throughout the network so that the majority of the nodes consume their power supply at relatively the same rate regardless of physical location. This leads to better maintainability of the system, such as replacing the batteries all at once rather than one by one, and maximizing the overall system performance by allowing the network to function at 100% capacity throughout most of its lifetime instead of having a steadily decreasing node population.

## 3. CONCLUSION

In summary, we showed that energy-efficient distributed dynamic diffusion routing is possible at very little overhead cost. The most significant outcome is the near optimal performance of e3D when compared to its ideal counterpart in which global knowledge is assumed between the network nodes. We therefore conclude that complex clustering techniques are not necessary in order to achieve good load and power usage balancing. Previous work suggested random clustering as a cheaper alternative to traditional clustering; however, random clustering cannot guarantee good performance according to our simulation results.